\documentclass[multphys,vecphys]{svmult}

\usepackage{makeidx}         
\usepackage{graphicx}        
\usepackage{multicol}        
\usepackage[bottom]{footmisc}

\makeindex             

\begin{document}

\title*{Evolution of the Luminosity-Metallicity-Stellar Mass  correlation in a hierarchical scenario}
\titlerunning{Luminosity-Metallicity-Stellar Mass  correlation}
\author{Mar\'{\i}a Emilia De Rossi\inst{1,2},
Patricia Beatriz Tissera\inst{1,2}\and
Cecilia Scannapieco\inst{1,2}}
\institute{Consejo Nacional de Investigaciones Cient\'{\i}ficas y T\'ecnicas, Argentina
\and Instituto de Astronom\'{\i}a y F\'{\i}sica del Espacio, Ciudad de Buenos Aires, Argentina  \texttt{derossi@iafe.uba.ar}}
%
%
\maketitle


\begin {abstract}
We study the evolution of the Stellar Mass-Metallicity Relation and
the Luminosity-Metallicity Relation by performing
numerical simulations in a cosmological framework.
We find that the slope and the zero point 
of the Luminosity-Metallicity Relation evolve in such
a way that, at a given metallicity, systems were 
 $\sim 3$ mag brighter at $z=3$ compared to galaxies in
the local universe, which is consistent with the
observational trend.  The local
Stellar Mass-Metallicity Relation 
shows also a good agreement  with recent observations.
We identify a characteristic stellar mass
 $M_{\rm c} \sim 10^{10.2} M_{\odot} h^{-1}$
at which the slope of the MMR decreases
for larger stellar masses.
Our results indicate
that $M_{\rm c}$
arises naturally
as a consequence of the hierarchical building up of the structure 
. 

\end {abstract}

\section{Results and discussion}
\label{sec:1}

The Luminosity-Metallicity Relation (LMR)
has been widely studied
during the last two decades.  In the local universe there is 
a clear correlation between the luminosity and the chemical
abundance of galaxies in such a way that brighter 
systems tend to be more metal rich
(e.g. \cite{Lamareille2004}). Futhermore,
recent observations have suggested that this correlation
evolves with redshift 
(e.g. \cite{Kobulnicky2003}). At a given luminosity
systems seem to be  less enriched in the past.
Because of the difficulties in obtaining stellar mass, 
most studies have used galaxy luminosity as a surrogate.
Recently, though, 
the relation between oxygen chemical abundance and stellar
mass (MMR) has been estimated
for local galaxies finding a well defined correlation
\cite{Tremonti2004}. 
In this work we study the MMR and the LMR by using
chemo-dynamical simulations in a cosmological scenario,
which allow us to follow the non-lineal growth of structure
together with the chemical enrichment of the interestellar
medium. 
 
We have performed chemo-dynamical simulations by employing
the chemical GADGET-2 \cite{Scannapieco2005}.
A $\Lambda$CDM universe
($\Omega =0.3, \Lambda =0.7, \Omega_{b} =0.04$  
and $H_{0} =100 h^{-1}$ km ${s}^{-1} {\rm Mpc}^{-1}, h=0.7$) 
was assumed. 
The simulated volume corresponds to a periodic cubic
box of a comoving 10 Mpc $h^{-1}$ side length.
We have considered two runs initially resolved with 
2 $\times \, 160^3$  and 2 $\times \, 80^3$  particles
(for details see De Rossi et al. 2006
in preparation). 

Our simulations predict a local LMR which
is in good agreement with the observational findings (e.g. 
\cite{Kobulnicky2003} among others). The slope
and the zero point of the LMR evolve with time 
in such a way that 
systems, at a given chemical abundance, are $\sim 3$ mag 
brighter at $z=3$ compared with local ones. The major
variations in chemical abundance are driven
by the faintest systems.
We have also analysed the MMR obtaining similar trends
to those encountered by \cite{Tremonti2004} 
for SDSS galaxies but with
a displacement of $\sim -0.25$ dex in the zero point.
This discrepancy may be due to the fact that the SDSS
explores only the central regions of galaxies leading
to an overestimation of the metal content of the systems.
We also observe an excess of metals 
at lower masses for the simulated MMR 
which may be probably associated to the
lack of strong energy feedback in our numerical model.

The metallicity of galactic systems tends to increase
with stellar mass in a non-linear way. 
We have determined a characteristic mass 
 $M_{\rm c} \sim 10^{10.2} M_{\odot} h^{-1}$
where the slope of the MMR decreases and
the relation starts to flatten for larger masses
\cite{Tissera2005}.
 It is important to note that this mass is independent of redshift
and has been previously mentioned in the literature 
as a characteristic mass
for galaxy evolution
\cite{Kauffmann2003}.
By examining the history of evolution of each individual
galactic system,
 we found that those with stellar masses $M_* > M_{\rm c}$ have experienced
important merger events at high redshift causing an acceleration
in the transformation of gas into stars.  At lower redshifts these 
systems are saturated of stars and hence, while
their stellar mass significantly increases during a merger
event, their metallicity remains almost the same.  On the other
hand  less
massive systems form their stars in a more passive way or by rich
gas mergers leading to a more tight correlation between 
stellar mass and metallicity.
Thus, the patterns of the MMR are closely related to
the hierarchical aggregation of structure in a LCDM universe
and hence it might be considered a fossil of this process.

MEDR thanks the LOC of this conference for their financial support.
This work has been partially supported by 
CONICET, Fundaci\'on Antorchas and LENAC.

%
%
%

%
%



\printindex
\end{document}